\documentclass{elsart}
\usepackage{epsfig}

\begin{document}
\runauthor{Cicero, Caesar and Vergil}

\begin{frontmatter} 

\title{ Theoretical estimates of cross sections for neutron-nucleus collisions }

\author{Tapan Mukhopadhyay\thanksref{X}}, 
\author{Joydev Lahiri\thanksref{Y}} and
\author{D.N. Basu\thanksref{Z}}

\address{Variable  Energy  Cyclotron  Centre, 1/AF Bidhan Nagar, Kolkata 700 064, India}

\thanks[X]{E-mail:tkm@veccal.ernet.in}
\thanks[Y]{E-mail:joy@veccal.ernet.in}
\thanks[Z]{E-mail:dnb@veccal.ernet.in}

\begin{abstract}

    We construct an analytical model derived from nuclear reaction theory and having a simple functional form to demonstrate the quantitative agreement with the measured cross sections for neutron induced reactions. The neutron-nucleus total, reaction and scattering cross sections, for energies ranging from 5 to 700 MeV and for several nuclei spanning a wide mass range are estimated. Systematics of neutron scattering cross sections on various materials for neutron energies upto several hundred MeV are important for ADSS applications. The reaction cross sections of neutrons are useful for determining the neutron induced fission yields in actinides and pre-actinides. The present model based on nuclear reaction theory provides good estimates of the total cross section for neutron induced reaction. 
\vspace{0.2cm}

\noindent
{\it PACS numbers}: 24.10.-i, 25.40.-h, 28.20.-v, 28.20.Cz, 28.20.Fc, 24.10.Ht, 25.60.Dz
\end{abstract}

\noindent
\begin{keyword}
n-N cross section; Ramsauer model; Optical model; ADSS; RIB.
\end{keyword}

\end{frontmatter}

\noindent
\section{Introduction}
\label{section1}

    The cross sections for neutron induced reactions by nuclei at very high energies upto several hundred MeV are required in a number of fields of study in basic science as well as many of applied nature such as radioactive ion beam (RIB) (Diamond 1999; Essabaa 2003) production or the accelerator-driven sub-critical systems (ADSS) and their neutronics (Boudard et al. 2000; Broeders and Broeders 2000; Demirkol et al. 2004; Kaplan et al. 2009; Yap\'yc\'y et al. 2007; Demirkol et al. 2008). The sub-critical reactor is driven critical by spallation neutrons produced by bombarding high energy proton beam with high current (greater than $\sim$ 10mA) (Rubbia et al. 1995) on a heavy element target. Such a system serves a dual purpose of energy multiplication and waste incineration of the long lived radioactive waste produced in reactors based on thermal neutron induced fission. Unlike the thermal neutron induced fission, the energy spectrum of spallation neutrons can reach up to several hundred MeV. In this context it is important to study the systematics of neutron absorption and scattering cross sections on various nuclei for neutron energies up to several hundred MeV. 

    Often, these cross sections are evaluated using phenomenological optical potentials and much effort has gone into defining global sets of parameter values for those optical potentials with which to estimate cross sections as yet unmeasured. In a recent study, Koning and Delaroche (Koning and Delaroche 2003) gave a detailed specification. In this context it would be utilitarian to study the systematics of neutron absorption and scattering cross sections on various nuclei well approximated by a simple convenient functional form (Deb and Amos 2004; Deb, Amos,and Karataglidis 2004). Indeed that is so and we show herein that the simple analytical model can be used to estimate these cross sections without recourse to optical potential calculations which are limited up to 150-200 MeV. 

\noindent
\section{Theoretical formalism}
\label{section2}

    Although the optical model calculations give reasonable account of all the data presented here, these calculations involve many partial waves and it is not easy to get a simple interpretation or intuitive picture of the basic processes involved. The basic picture of the nuclear Ramsauer model is that the forward scattering amplitude for a neutron incident on a nucleus is given by 

\begin{equation}
 f(0^o) = \frac{i}{2k}\sum_{l=0}^L (2l+1)(1-e^{2i\delta_l})
\label{seqn1}
\end{equation}
\noindent
where the complex phase shift $\delta_l$ may be considered as independent of angular momenta $l$ of the $l$th partial wave (Lawson 1953), so that the above expression, after summing over $l$ upto a maximum value $L=kR_{ch}$ (Angeli and Csikai 1970; Angeli and Csikai 1971; Angeli, Csikai and Nagy 1974), reduces to 

\begin{equation}
 f(0^o) = ik(R_{ch}+\Lambda)^2(1-\alpha e^{i\beta})/2
\label{seqn2}
\end{equation}
\noindent
where $\Lambda=1/k=\hbar/\sqrt{2mE}$, $E$ is the incident neutron energy in the center of mass system, $m$ is the reduced mass of the neutron-nucleus system and $R_{ch}$ is the channel radius beyond which partial waves do not contribute. The quantity $e^{2i\delta_l}$ is replaced by $\alpha e^{i\beta}$ with $\beta$ being two times the real part of the phase shift and $\alpha$ accounting for the attenuation or loss of flux (arising out of imaginary part of the phase shift). From partial wave analysis of scattering theory, we know the standard expressions for scattering $\sigma_{sc}$ and reaction $\sigma_r$ cross sections as

\begin{equation}
\sigma_{sc}=\frac{\pi}{k^2} \Sigma_l~(2l+1)|1-\eta_l|^2,~~~~\sigma_r=\frac{\pi}{k^2} \Sigma_l~(2l+1)[1-|\eta_l|^2]
\label{seqn3}
\end{equation}
\noindent
where the quantity $\eta_l = e^{2i\delta_l}$. With the assumption that the phase shift $\delta_l$ is independent of $l$ and the summation over partial waves $l$ is upto $kR_{ch}$ only, it follows that

\begin{eqnarray}
\sigma_{sc}=&&\pi(R_{ch}+\Lambda)^2(1+\alpha^2-2\alpha \cos\beta),  \nonumber\\
\sigma_r=&&\pi(R_{ch}+\Lambda)^2(1-\alpha^2),  \nonumber\\ 
\sigma_{tot}=&&\sigma_{sc} + \sigma_r = 2\pi(R_{ch}+\Lambda)^2(1-\alpha \cos\beta)
\label{seqn4}
\end{eqnarray}
\noindent
where $\beta=2 {\rm Re}\delta_l = 2 {\rm Re}\delta $, $\alpha = e^{-2 {\rm Im}\delta_l} = e^{-2 {\rm Im}\delta}$ and summing over $l$ from 0 to $kR_{ch}$ yields $\Sigma_l~(2l+1) = (kR_{ch}+1)^2 = k^2 (R_{ch}+\Lambda)^2$. 

    The optical potential for nuclear n-N interaction can be written as $-V-iW$ with $V$ and $W$ as positive quantities, and contains no Coulomb interaction. The phase shift $\delta$ in a WKB approximation is $[\int K'dr-\int k'dr]$ and the real part of it to a zeroth order approximation (Mohr 1957) for a square well with radius $R$ is $(K-k) R$ where $K$ is the real part of $K'$ and $k'=k$ is real due to absence of potential. The real wave numbers inside and outside the nucleus are, therefore, given by $K^2=2m(E+V)/\hbar^2$ and $k^2=2mE/\hbar^2$ respectively. Hence $\beta$ is determined by the real potential $V$,

\begin{equation}
 \beta=2(K-k) R = 2 \frac{(2m)^{\frac{1}{2}}}{\hbar} [\sqrt{E+V}- \sqrt{E}] R
\label{seqn5}
\end{equation}
\noindent
whereas the attenuation factor $\alpha$ is determined primarily by the imaginary potential $W$, 

\begin{equation}
 \alpha=e^{-\bar R/\lambda}=e^{-2mW\bar R/\hbar^2K}
\label{seqn6}
\end{equation}
\noindent
where $\lambda$ is the mean free path of the neutron inside the nucleus. The average chord length $\bar R$ of a neutron passing through a nucleus can be derived as

\begin{equation}
 \bar R=\int_0^R 2 \sqrt{R^2-x^2} (I 2\pi x dx) \Big / \int_0^R I 2\pi x dx =\frac{4}{3}R
\label{seqn7}
\end{equation}
\noindent
where $I$ is the neutron flux that is the number of neutrons incident per unit area. Since $R\propto A^{\frac{1}{3}}$ ($R\sim r_0 A^{\frac{1}{3}}$), the above arguments imply that 

\begin{equation}
 \beta=\beta_0 A^{\frac{1}{3}}[\sqrt{E+V}- \sqrt{E}]
\label{seqn8}
\end{equation}
\noindent
where $\beta_0=\frac{2r_0(2m)^{\frac{1}{2}}}{\hbar}$ whose value is approximately 0.6, and the attenuation factor which is much less than unity but increases with energy (which is obvious from its expression) is given by 

\begin{equation}
 \alpha=\exp[{-\alpha_0 r_0 A^{\frac{1}{3}} W/\sqrt{E+V}}]
\label{seqn9}
\end{equation}
\noindent
where $\alpha_0=\frac{4(2m)^{\frac{1}{2}}}{3\hbar}$ whose value turns out to be 0.2929 and $r_0$ is the nuclear radius parameter. The first term $V_A=V_0+ V_1(1-2Z/A)+V_2/A$ of the real potential $V=V_A + V_E\sqrt{E}$ contains both the isoscalar and the isovector (Lane 1962; Satchler 1983) components of the optical potential (Gould 1986; Anderson and Grimes 1990) where $Z$ is the atomic number of the target nucleus, whereas the second term accounts for its energy dependence. The imaginary potential $W$ is taken as $W=W_0 + W_E\sqrt{E+V}$ since the total kinetic energy of the neutron inside the nucleus with attractive potential well of depth $V$ is $E+V$. As the magnitude of the real part of the optical potential decreases with energy while the same for imaginary part increases, this implies that $V_E$ is negative whereas the $W_E$ is positive.

    It is worthwhile to mention that in all the previously published papers, the expression for the phase shift $\beta$ has been explained as caused by refraction in a sphere of radius $R$ with refractive index $n$ like Peterson (Peterson 1962). We point out here that the derivation of ref.(Peterson 1962) is not profound and provided above an alternative, theoretically concrete, derivation for the phase shift using WKB method of quantum mechanics as applied to scattering by a spherical potential well. The drawback of the derivation of ref.(Peterson 1962) is that the neutron (although massive) is treated like a photon and as its velocity inside nucleus and vacuum are proportional to $\sqrt{E+V}$ and $\sqrt{E}$, respectively, it would result in bending of the ray (as in optics) away from the normal inside nucleus (in fig.16 of ref.(Peterson 1962) it is shown just the opposite) where velocity is more. This would lead to the existence of the critical angle $sin^{-1}\sqrt{E/(E+V)}$ beyond which there is no transmission (even in an attractive nuclear potential) and a refractive index less than vacuum for the nuclear medium which are physically unacceptable. Even then, if one sticks to Peterson's assumption of a light ray, the average chord length inside nucleus, with ray bending away from the normal, turns out to be less than our result of $4R/3$ as opposed to greater than $4R/3$ as derived in ref.(Peterson 1962). However, these results go over to our result of $4R/3$ for refractive index equal to one, that is, reach our result asymptotically at energies higher than magnitude of the real part of the nuclear potential. 

\noindent
\section{ The analytical model calculations }
\label{section3}

    The Ramsauer model can be fitted to the experimental neutron total cross sections using Eq.(4). The radius of the nuclear potential is given by $R=r_0 A^{\frac{1}{3}}$ whereas the channel radius can be parametrized (Gowda and Ganesan 2006) as $R_{ch}=r_0 A^{\frac{1}{3}}+r_A \sqrt{E}+r_2$ with $r_A= r_{10}~{\rm ln}A+ r_{11}/{\rm ln}A$. The Ramsauer model fits yield $r_{10} =-22.98\times 10^{-3}$, $r_{11} = 10.27\times 10^{-2}$, $r_2 = 23.22\times 10^{-2}$, $V_0 = 46.51$, $V_1 = 6.74$, $V_2 = -117.52$, $V_E = -3.22$ and $\beta_0 = 0.5928$. These values are very close to or within the limit of the parameter values obtained in our earlier work (Mukhopadhyay, Lahiri and Basu 2010). The value of $\alpha_0$ is kept fixed at 0.2929 and the non linear least square fits yield the values for the imaginary potential $W_0=5.293$ MeV and its energy dependence $W_E=33.88\times 10^{-2}$. The nuclear radius parameter $r_0$ is also fitted reasonably well to $1.378~A^\gamma$ fm which means that the nuclear potential radius $R=r_0^\prime A^{\frac{1}{3}+\gamma}$ where $\gamma=7.93\times 10^{-3}$ is a very small number (needed for fine tuning) compared to $\frac{1}{3}$. These extracted model parameters provide global fits to the neutron total cross sections spanning quite a large number of target nuclei. In our earlier work (Mukhopadhyay, Lahiri and Basu 2010) the effect of the imaginary potential was included in the parameter $\alpha$ in somewhat {\it ad-hoc} manner lacking justification of it to be weakly mass dependent. The present work is an improvement over our earlier work where the imaginary part of the nuclear potential is now treated appropriately with explicit appearance of the imaginary part of the nuclear potential.

\noindent
\section{ Results }
\label{section4}

    Each calculation is performed at neutron incident energy intervals of 1 MeV and for various elements. In Figs. 1-15, the variations of the total cross section ($\sigma_{tot}$), the scattering cross section ($\sigma_{sc}$) and the reaction cross section ($\sigma_r$) with incident neutron energy are plotted for $^{238}$U, $^{232}$Th, $^{209}$Bi, $^{208}$Pb, $^{197}$Au, $^{186}$W, $^{184}$W, $^{182}$W, $^{181}$Ta, $^{93}$Nb, $^{90}$Zr, $^{59}$Co, $^{55}$Mn, $^{40}$Ca and $^{31}$P target nuclei. The continuous lines represent the total cross sections, the dashed lines represent the scattering cross sections and the dotted lines represent the reaction cross sections while the hollow circles represent the experimental data (Finlay et al. 1993; Abfalterer et al. 2001; Dietrich et al. 2003) for $\sigma_{tot}$ which show good agreement with the measured cross sections. In Fig. 16 estimates of these cross sections are plotted for $^{239}$Pu target. 

\noindent
\section{ Summary and conclusion }
\label{section5}

    In summary, we constructed an analytical model, justified it from the optical model and nuclear reaction theory approach and applied it to derive the systematics and performed calculations of neutron-nucleus total cross section ($\sigma_{tot}$), the scattering cross section ($\sigma_{sc}$) and the reaction cross section ($\sigma_r$) and predicted these cross sections for heavier actinides. The extracted parameters for the present analytical model provide global fits to the neutron total cross sections spanning quite a large number of nuclei. We conclude that the present estimates of neutron scattering cross sections are very important for the reactor physics calculations Diamond 1999; Essabaa 2003) production or the accelerator-driven sub-critical systems (ADSS) and their neutronics (Boudard et al. 2000; Broeders and Broeders 2000; Demirkol et al. 2004; Kaplan et al. 2009; Yap\'yc\'y et al. 2007; Demirkol et al. 2008; Rubbia et al. 1995) of the ADSS applications. Moreover, the neutron-nucleus reaction cross sections are very useful for the theoretical calculations of Radioactive Ion Beam (Diamond 1999; Essabaa 2003) production. These can also be used for performing Hauser-Feshbach (Hauser and Feshbach 1952) calculations with Monte-Carlo simulations (Pace2 code 1984) to estimate the cross sections for neutron induced fission, evaporation residues or evaporation neutron multiplicities and for comparison of the photon (Mukhopadhyay and Basu 2007; Mukhopadhyay and Basu 2009) versus neutron induced fission as well and may serve as inputs to intranuclear cascade codes such as MCNPX package. 

\noindent
{\bf References}

Abfalterer, W.P., et al., 2001. Measurement of neutron total cross sections up to 560 MeV., Phys. Rev {\bf C 63}, 044608.

Anderson, J.D., Grimes, S.M., 1990. Nuclear Ramsauer effect and the isovector potential., Phys. Rev. {\bf C 41}, 2904.

Angeli, I., Csikai, J., 1970. Total neutron cross sections and the nuclear Ramsauer effect., Nucl. Phys. {\bf A 158}, 389.

Angeli, I., Csikai, J., 1971. Total neutron cross sections and the nuclear Ramsauer effect (II). E$_n$ = 0.5-42 MeV., Nucl. Phys. {\bf A 170}, 577.

Angeli, I., Csikai, J., Nagy, P., 1974. Semiclassical description of fast-neutron cross sections., Nucl. Sci. Eng. {\bf 55} , 418.

Boudard, A., et al., 2000. Spallation study with proton beams around 1 GeV : neutron production., Nucl. Phys. {\bf A 663}, 1061.

Broeders,  C. H. M., Broeders, I., 2000. Neutron physics analyses of accelerator- driven subcritical assemblies., Nucl. Eng. Des. {\bf 202 (2)} 209.

Deb, P.K., Amos, K., 2004. Simple function forms and nucleon-nucleus total cross sections., Phys. Rev. {\bf C 69}, 064608.

Deb, P.K., Amos, K., Karataglidis, S., 2004.  Simple functional form for the n+$^{208}$Pb total cross section between 5 and 600 MeV., Phys. Rev. {\bf C 70}, 057601.

Demirkol, I. et al., 2004. The neutron production cross sections for Pb, Bi and Au targets and neutron multiplicity for nuclear spallation reaction induced by 20-1600 MeV protons., Nucl. Sci. Eng. {\bf 147}, 83.

Demirkol, I., et al., 2008. Neutron multiplicity with 1.0 and 1.2GeV Proton-Induced spallation reactions on thin targets., Chinese Jour. Physics {\bf 46 (2)}, 124.

Diamond, W.T., 1999. A radioactive ion beam facility using photfission., Nucl. Instr. and Meth. {\bf A 432}, 471.

Dietrich, F.S., et al., 2003. Importance of isovector effects in reproducing neutron total cross section differences in the W isotopes., Phys. Rev. {\bf C 67}, 044606. 

Essabaa, S., et al., 2003. Photo-fission for the production of radioactive beams ALTO project., Nucl. Instr. and Meth. {\bf B 204}, 780. 

Finlay, R.W., et al., 1993. Neutron total cross sections at intermediate energies., Phys. Rev {\bf C 47}, 237.

Gould, C.R., et al., 1986. Spin-spin potentials in $^{27}$Al$_{pol}$ + n$_{pol}$ and the nuclear Ramsauer effect.,  Phys. Rev. Lett. {\bf 57}, 2371.

Gowda, R.S., Ganesan, S., 2006. Investigation of the Ramsauer model for the prediction of neutron cross sections below 120 MeV., Nucl. Sci. and Eng. {\bf 152}, 23.

Hauser, W., Feshbach, H., 1952. The inelastic scattering of neutrons., Phys. Rev. {\bf 87}, 366.

Kaplan, A., et al., 2009. Spallation neutron emission spectra in medium and heavy target nuclei by a proton beam up to 140 MeV energy., Applied Radia. and Isotopes {\bf 67}, 570.

Koning, A.J., Delaroche, J.P., 2003. Local and global nucleon optical models from 1 keV to 200 MeV., Nucl. Phys. {\bf A 713}, 231.

Lane, A.M., 1962. Isobaric spin dependence of the optical potential and quasi-elastic (p,n) reactions., Nucl. Phys. {\bf 35}, 676.

Lawson, J.D., 1953. A Diffraction effect illustrating the transparency of nuclei to high energy neutrons., Phil. Mag. {\bf 44}, 102.

Mohr, C.B.O., 1957. The WKB method for a complex potential., Austr. Jour. Phys. {\bf 10}, 110.

Mukhopadhyay, T., Basu, D.N., 2007. Photonuclear reactions of actinide and pre-actinide nuclei at intermediate energies., Phys. Rev. {\bf C 76}, 064610.

Mukhopadhyay, T., Basu, D.N., 2009. $\gamma$ induced multiparticle emissions of medium mass nuclei at intermediate energies., Phys. Rev. {\bf C 79}, 017602. 

Mukhopadhyay, T., Lahiri, J., Basu, D.N., 2010. Cross sections of neutron-induced reactions., Phys. Rev. {\bf C 82}, 044613 (2010). 

PACE2, 1984. Projection Angular momentum Coupled Evaporation code.

Peterson, J.M., 1962. Neutron giant resonances - Nuclear Ramsauer effect., Phys. Rev. {\bf 125}, 955.

Rubbia, C., et. al., 1995. Conceptual design of a fast Neutron Operated High power Energy Amplifier., CERN Rep, CERN/AT/95-44 (ET), Geneva, 29 Sep.

Satchler, G.R., 1983. Int. series of monographs on Physics, Oxford University Press, Direct Nuclear reactions, p471. 

Yap\'yc\'y, H., et al., 2007. Investigation of the properties of the nuclei used on the new generation reactor technology systems., Ann. Nucl. Energy, {\bf 34 (5)}, 374.

\pagebreak

\begin{figure}[h]
\eject\centerline{\epsfig{file=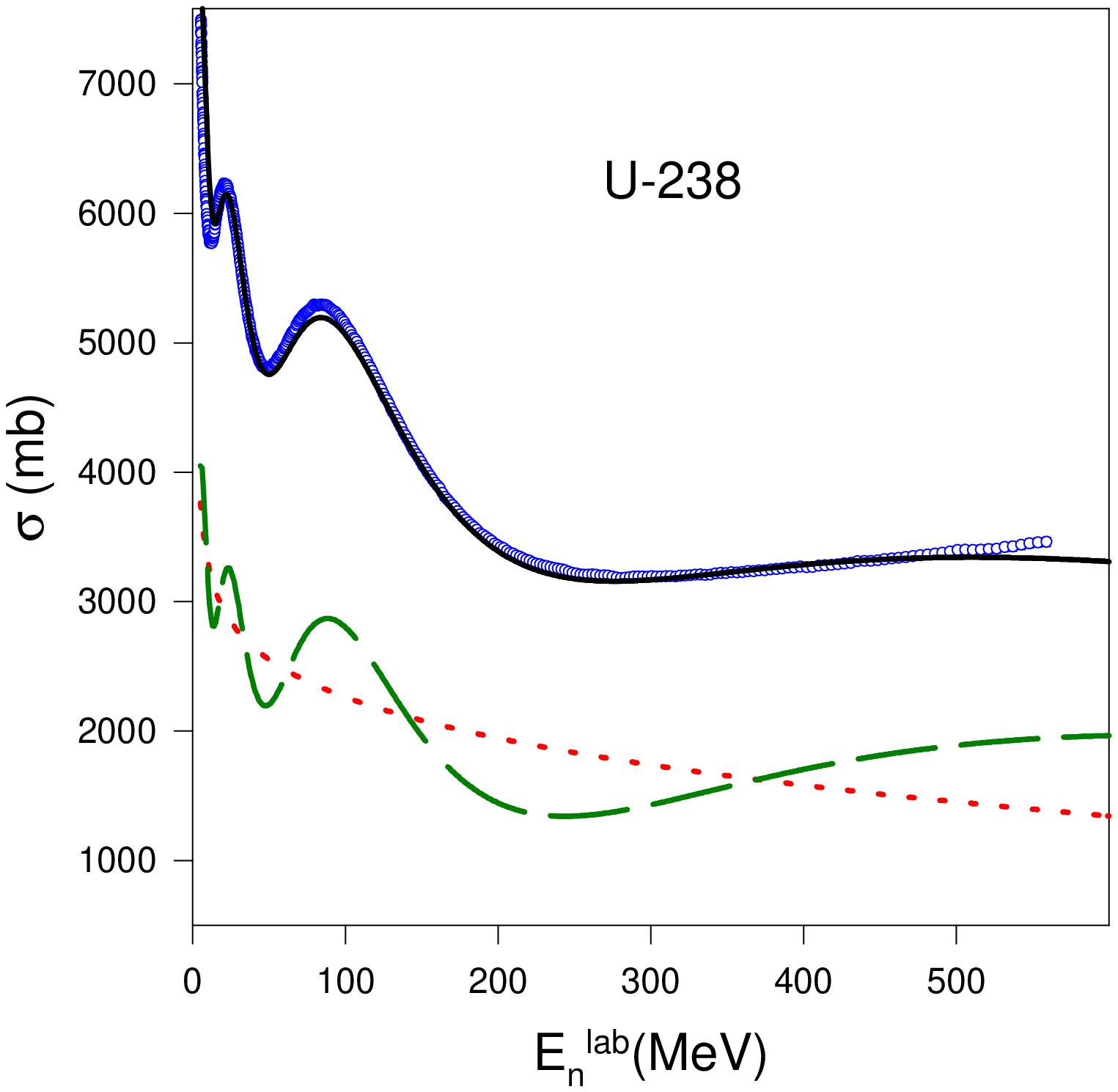,height=10cm,width=10cm}}
\caption{}
\label{fig1}
\end{figure}
\pagebreak

\begin{figure}[h]
\eject\centerline{\epsfig{file=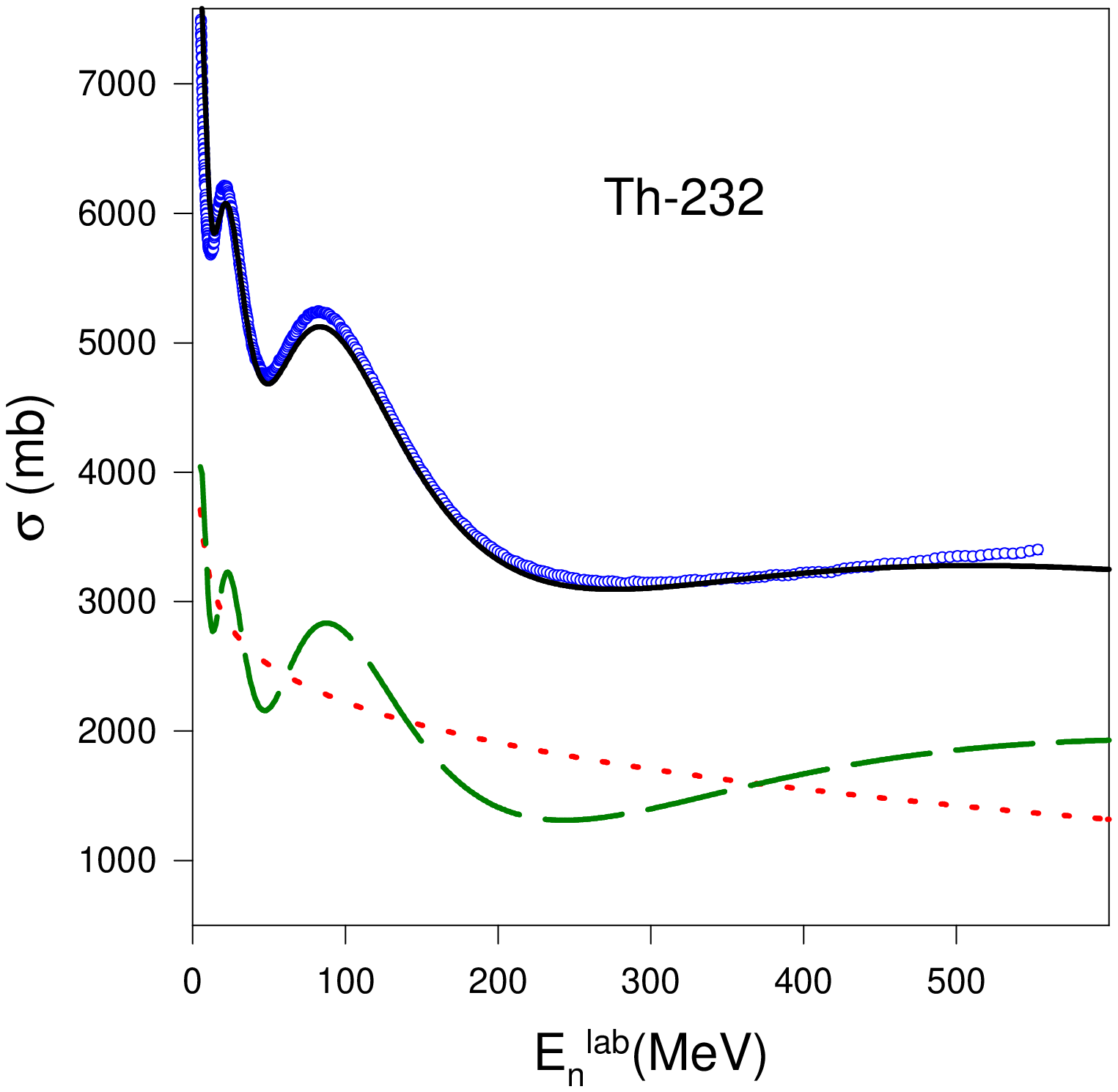,height=10cm,width=10cm}}
\caption{}
\label{fig2}
\end{figure}
\pagebreak

\begin{figure}[h]
\eject\centerline{\epsfig{file=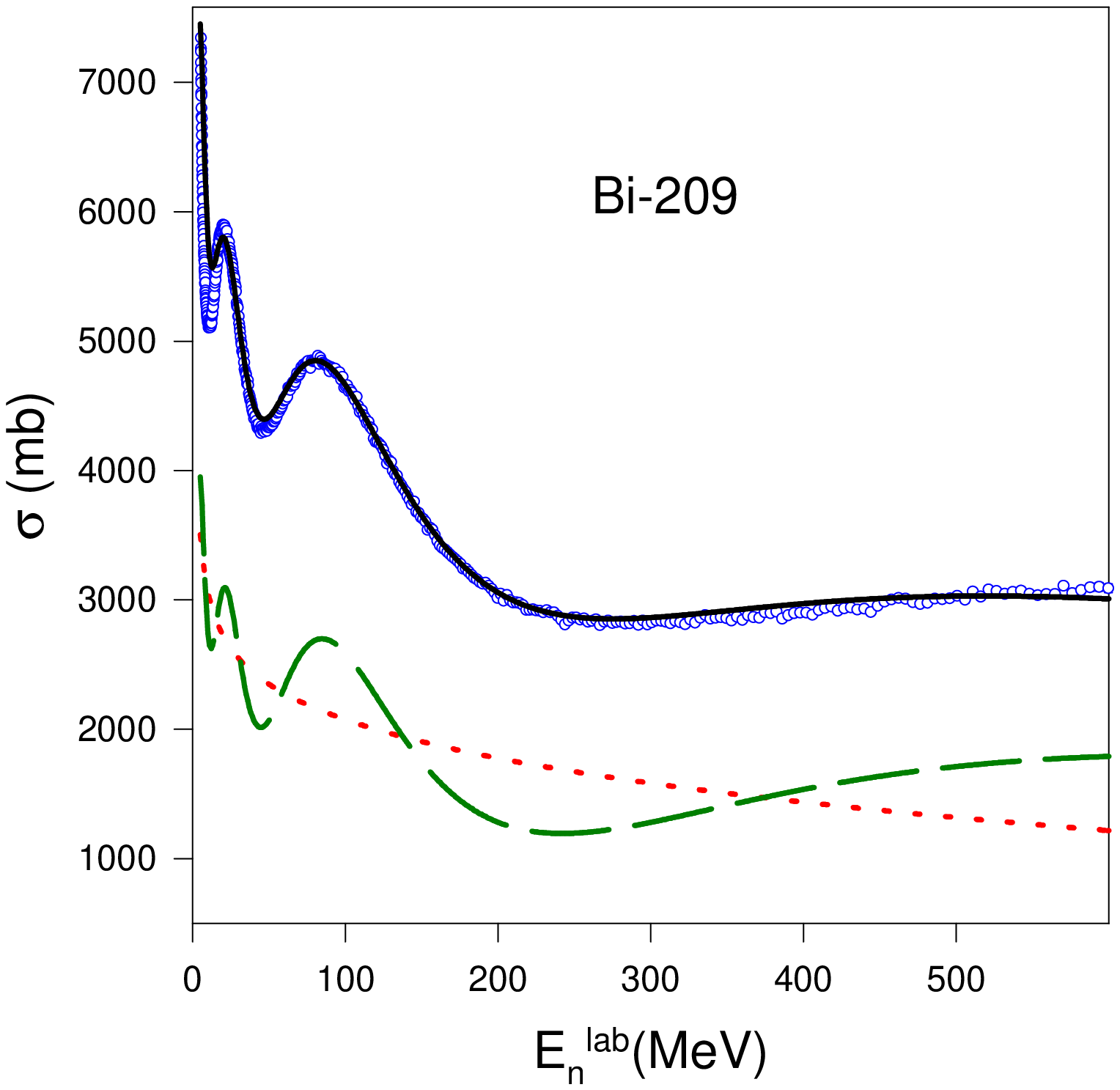,height=10cm,width=10cm}}
\caption{}
\label{fig3}
\end{figure}
\pagebreak

\begin{figure}[h]
\eject\centerline{\epsfig{file=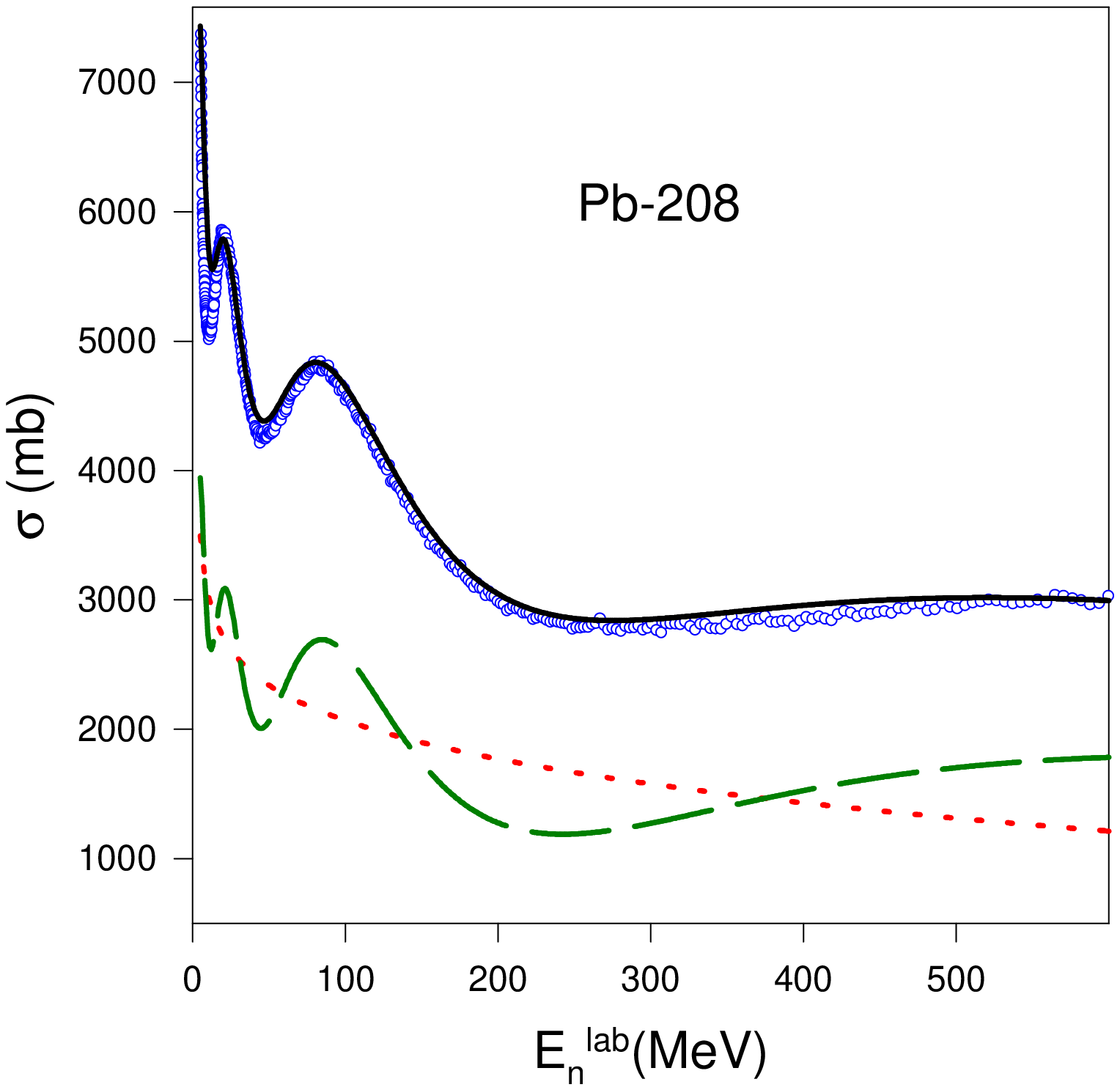,height=10cm,width=10cm}}
\caption{}
\label{fig4}
\end{figure}
\pagebreak

\begin{figure}[h]
\eject\centerline{\epsfig{file=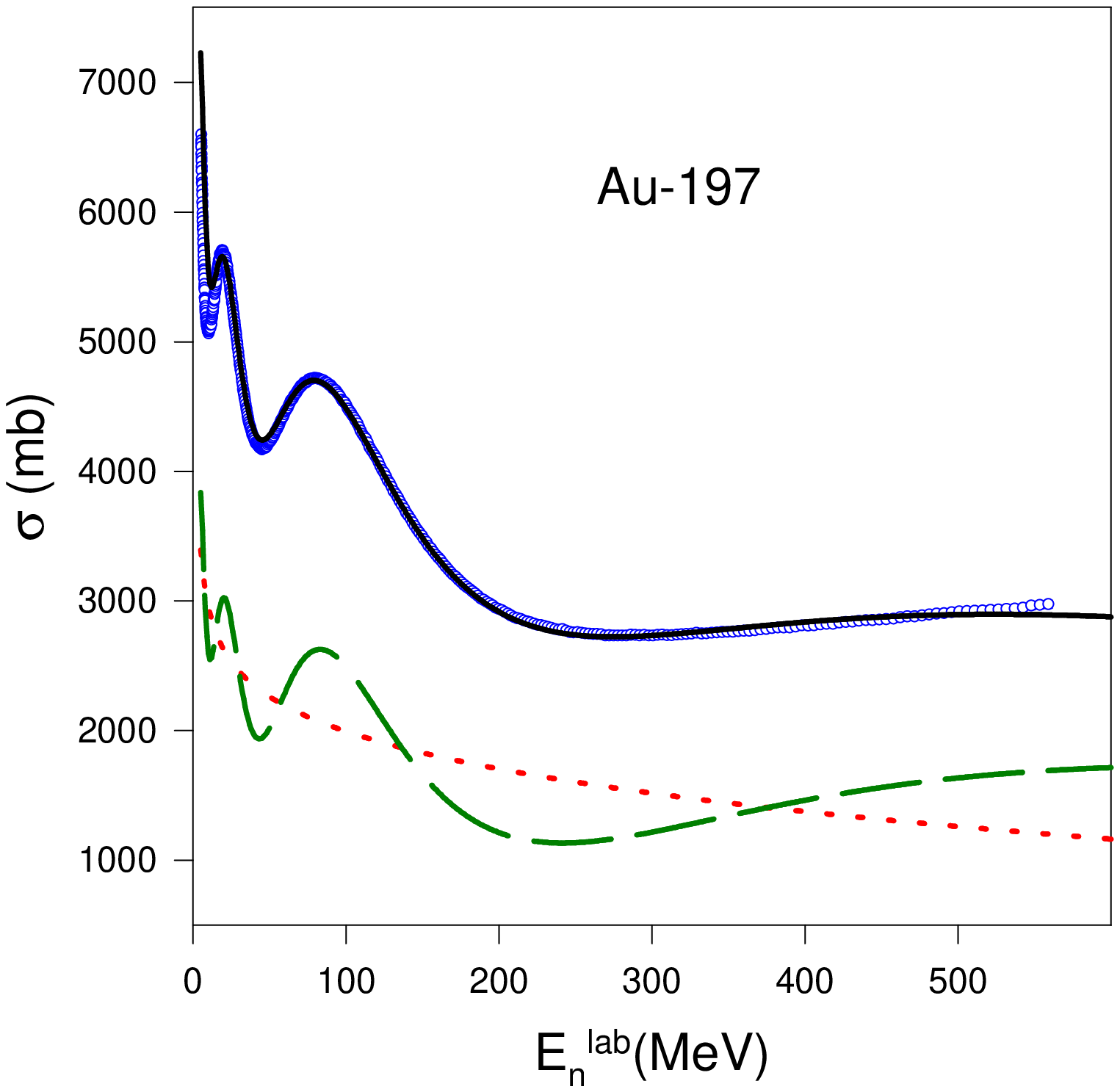,height=10cm,width=10cm}}
\caption{}
\label{fig5}
\end{figure}
\pagebreak

\begin{figure}[h]
\eject\centerline{\epsfig{file=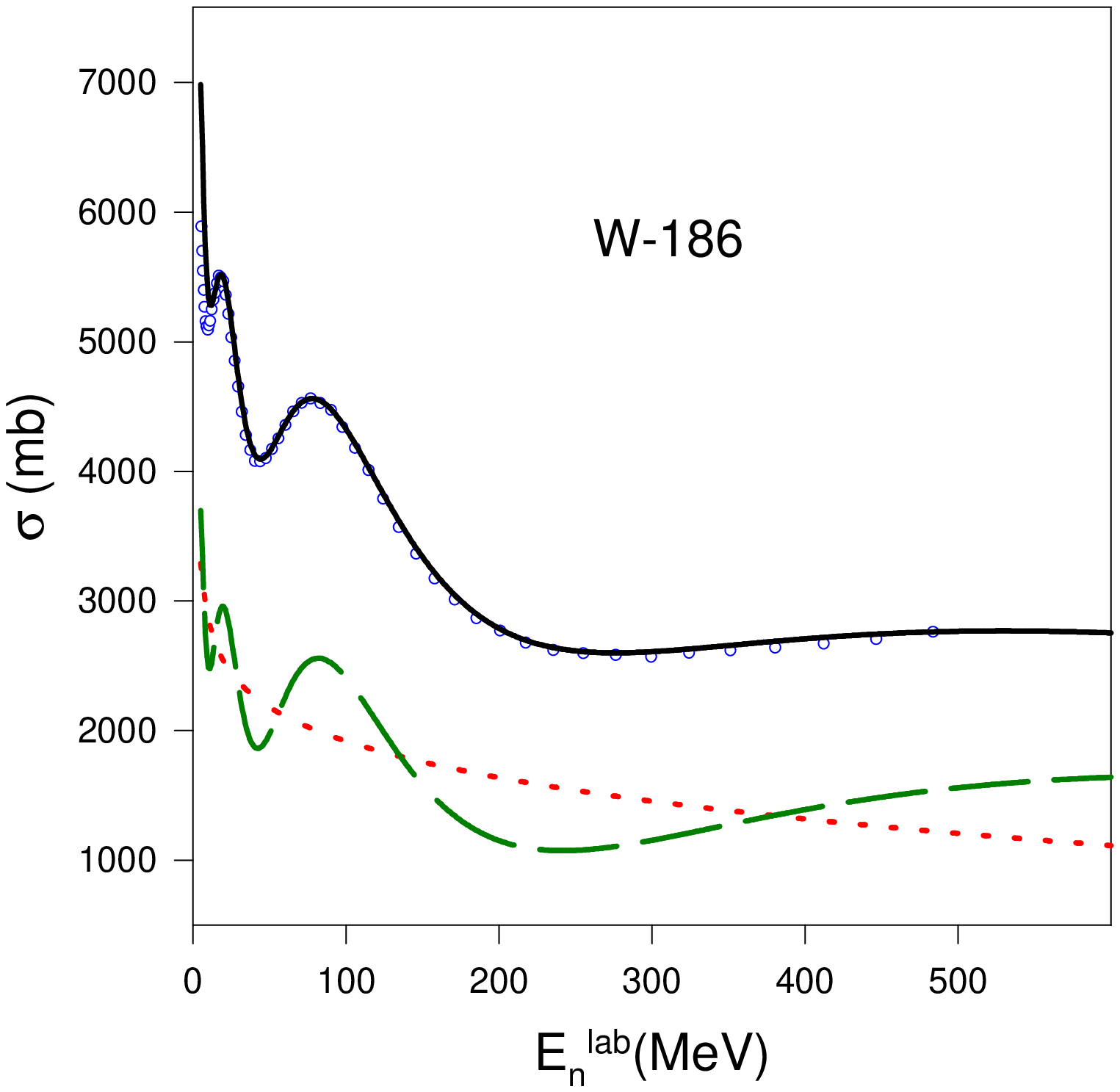,height=10cm,width=10cm}}
\caption{}
\label{fig6}
\end{figure}
\pagebreak

\begin{figure}[h]
\eject\centerline{\epsfig{file=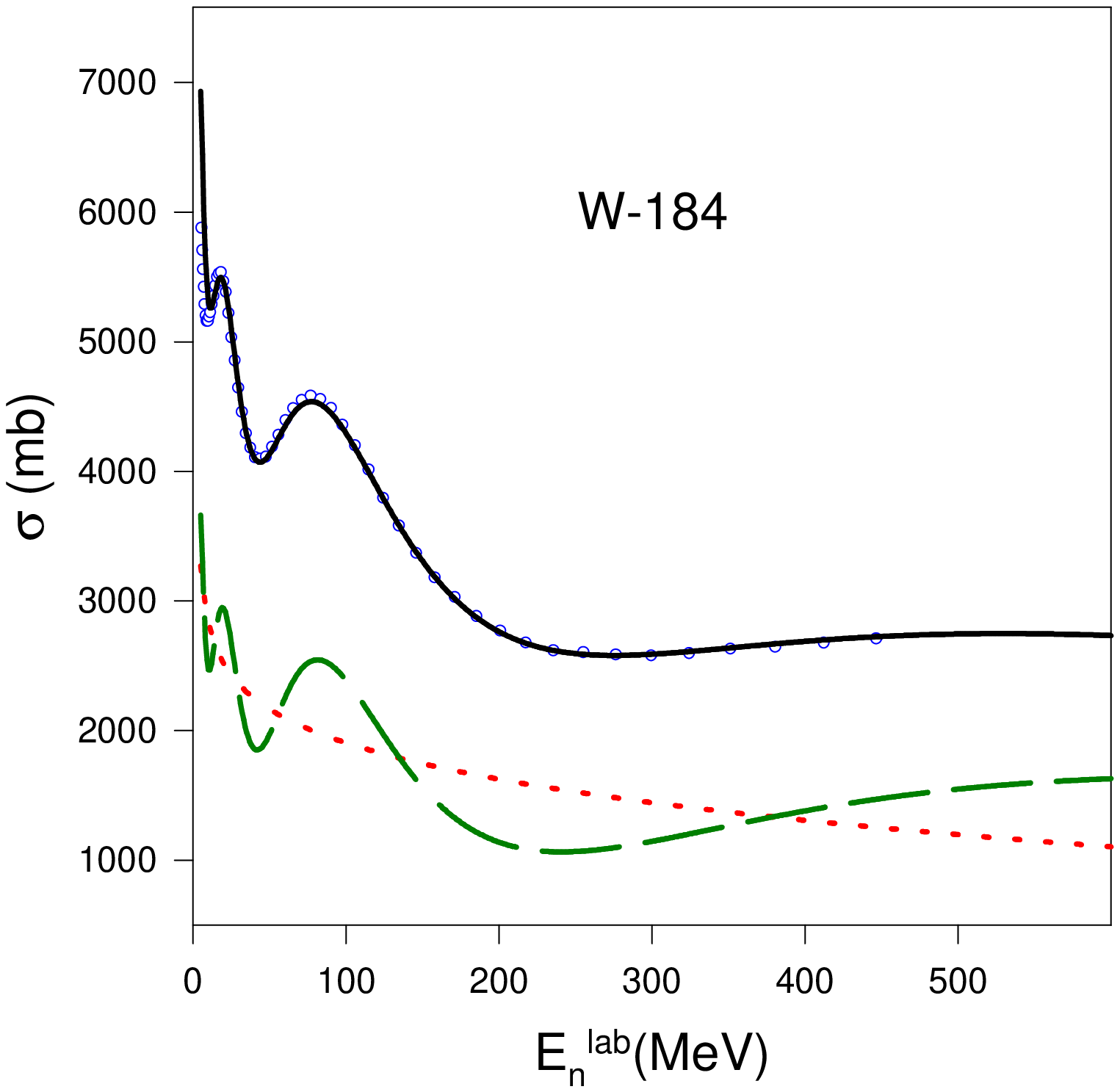,height=10cm,width=10cm}}
\caption{}
\label{fig7}
\end{figure}
\pagebreak

\begin{figure}[h]
\eject\centerline{\epsfig{file=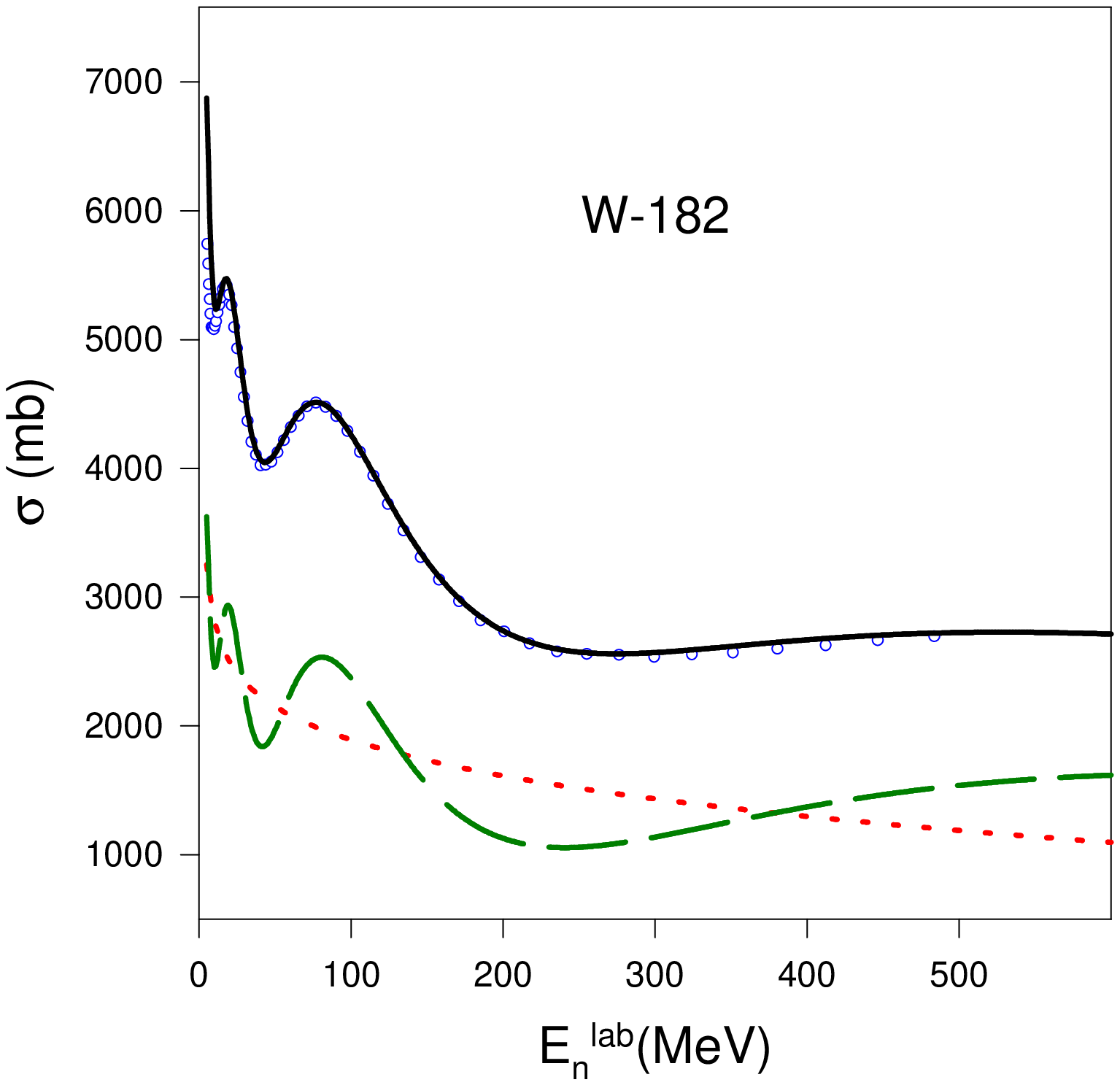,height=10cm,width=10cm}}
\caption{}
\label{fig8}
\end{figure}
\pagebreak

\begin{figure}[h]
\eject\centerline{\epsfig{file=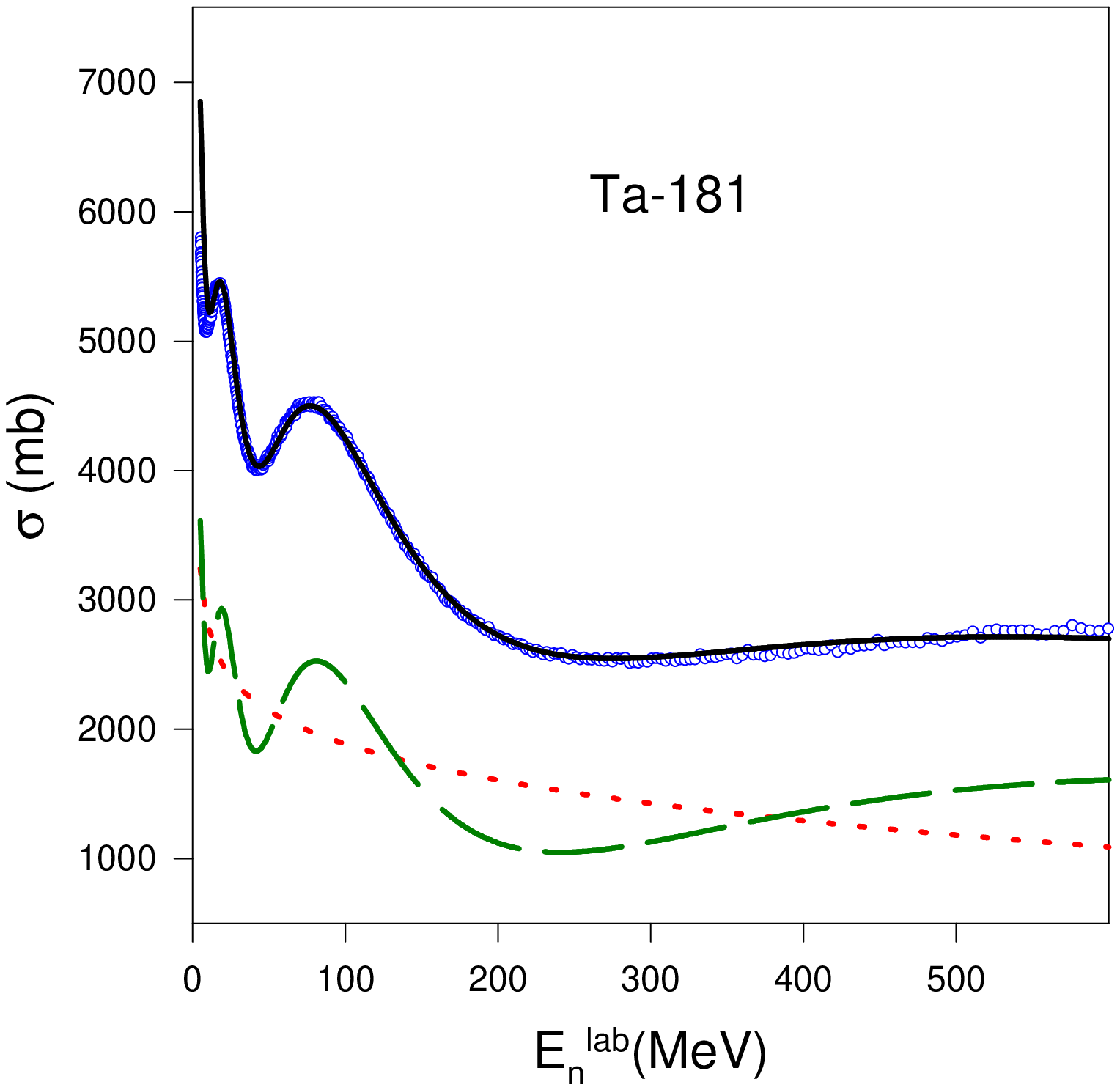,height=10cm,width=10cm}}
\caption{}
\label{fig9}
\end{figure}
\pagebreak

\begin{figure}[h]
\eject\centerline{\epsfig{file=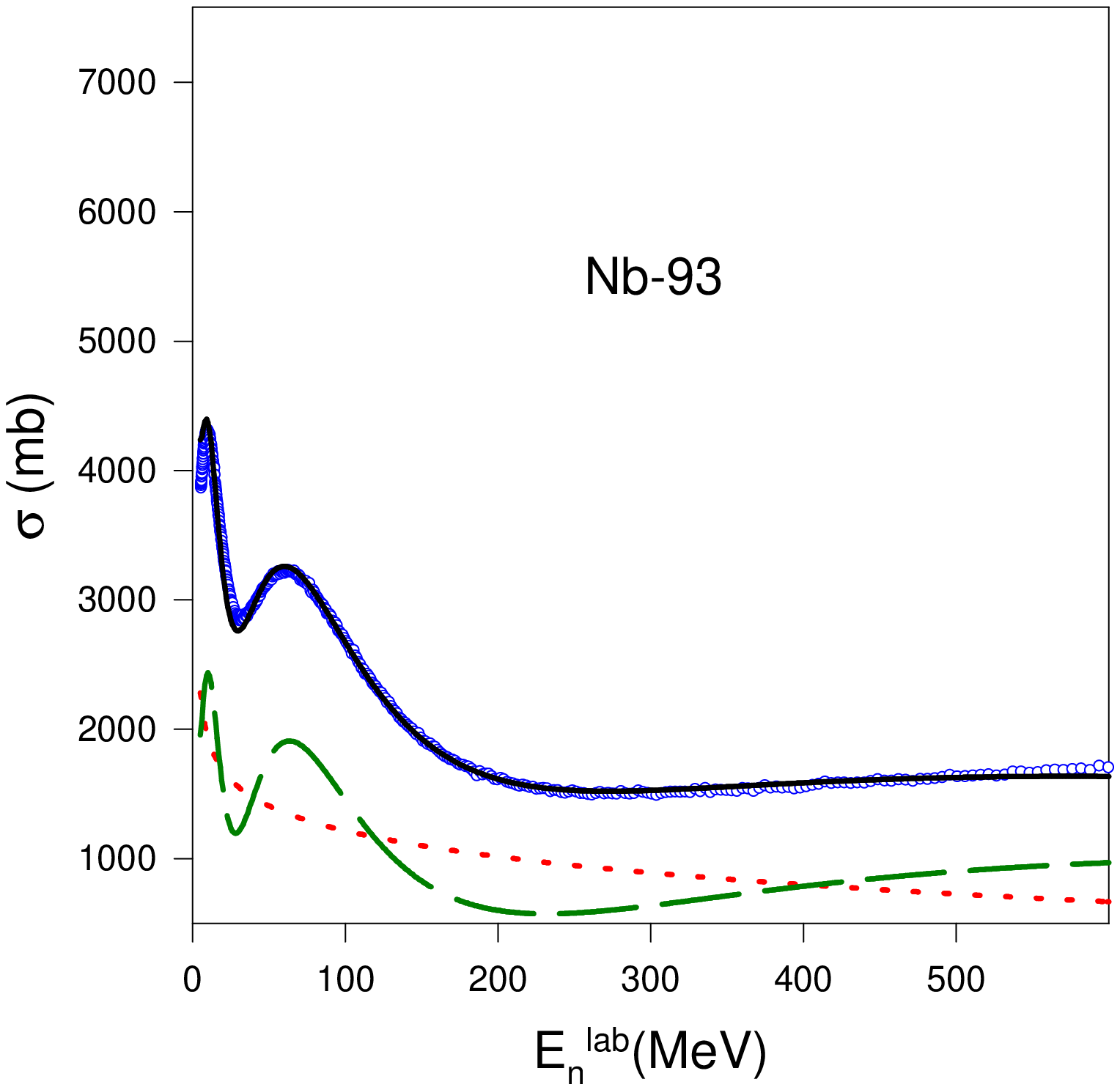,height=10cm,width=10cm}}
\caption{}
\label{fig10}
\end{figure}
\pagebreak

\begin{figure}[h]
\eject\centerline{\epsfig{file=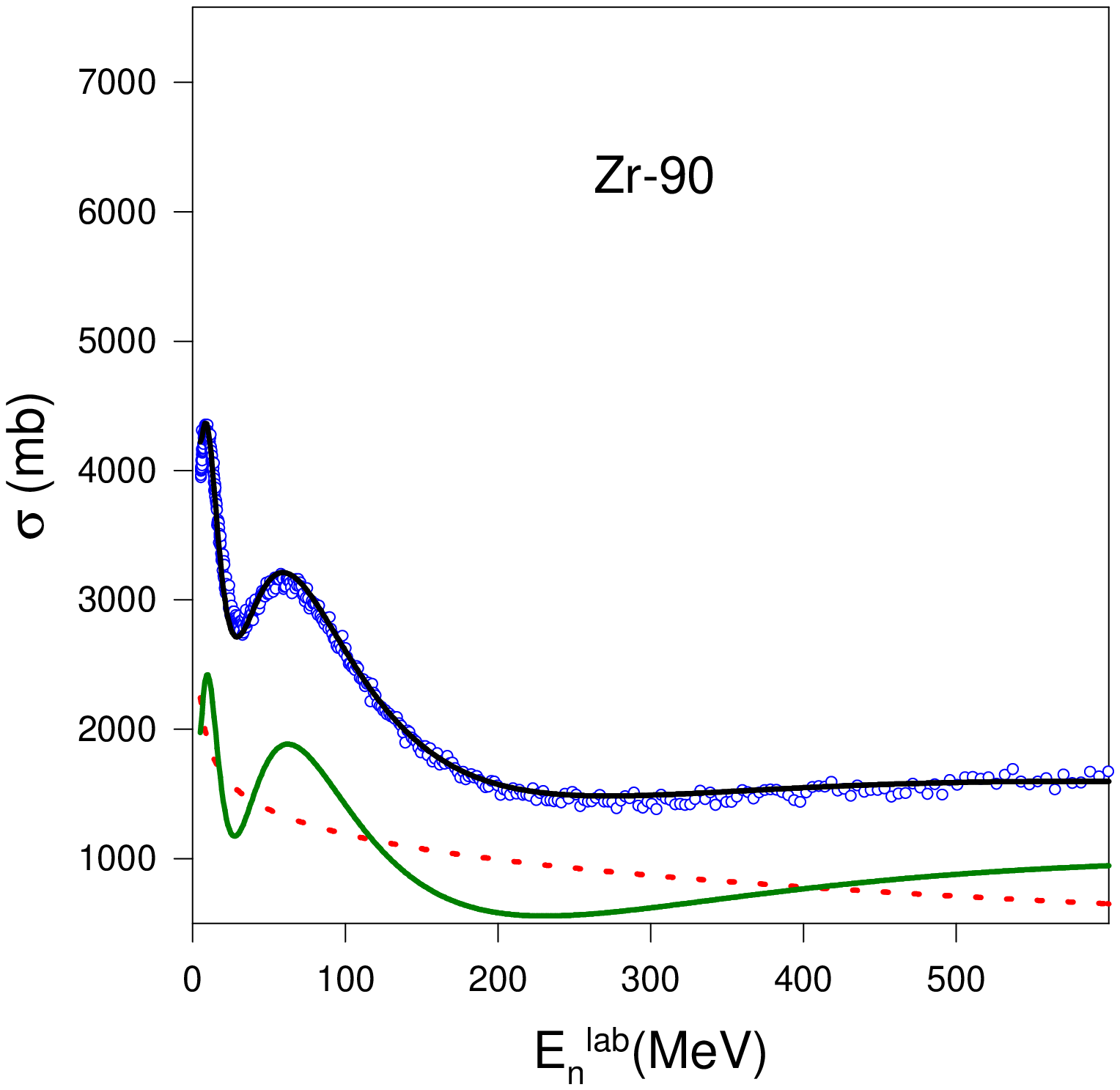,height=10cm,width=10cm}}
\caption{}
\label{fig11}
\end{figure}
\pagebreak

\begin{figure}[h]
\eject\centerline{\epsfig{file=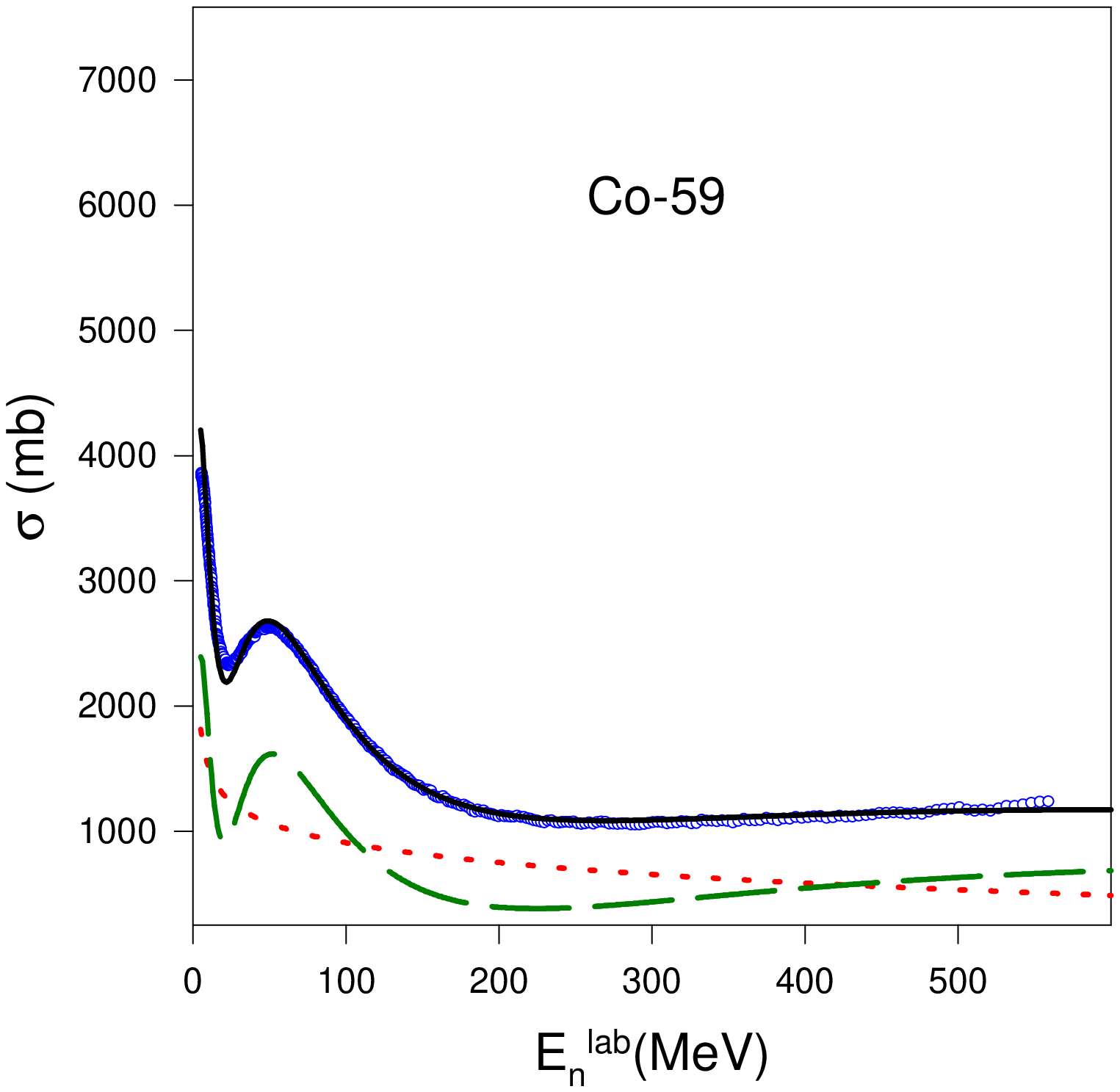,height=10cm,width=10cm}}
\caption{}
\label{fig12}
\end{figure}
\pagebreak

\begin{figure}[h]
\eject\centerline{\epsfig{file=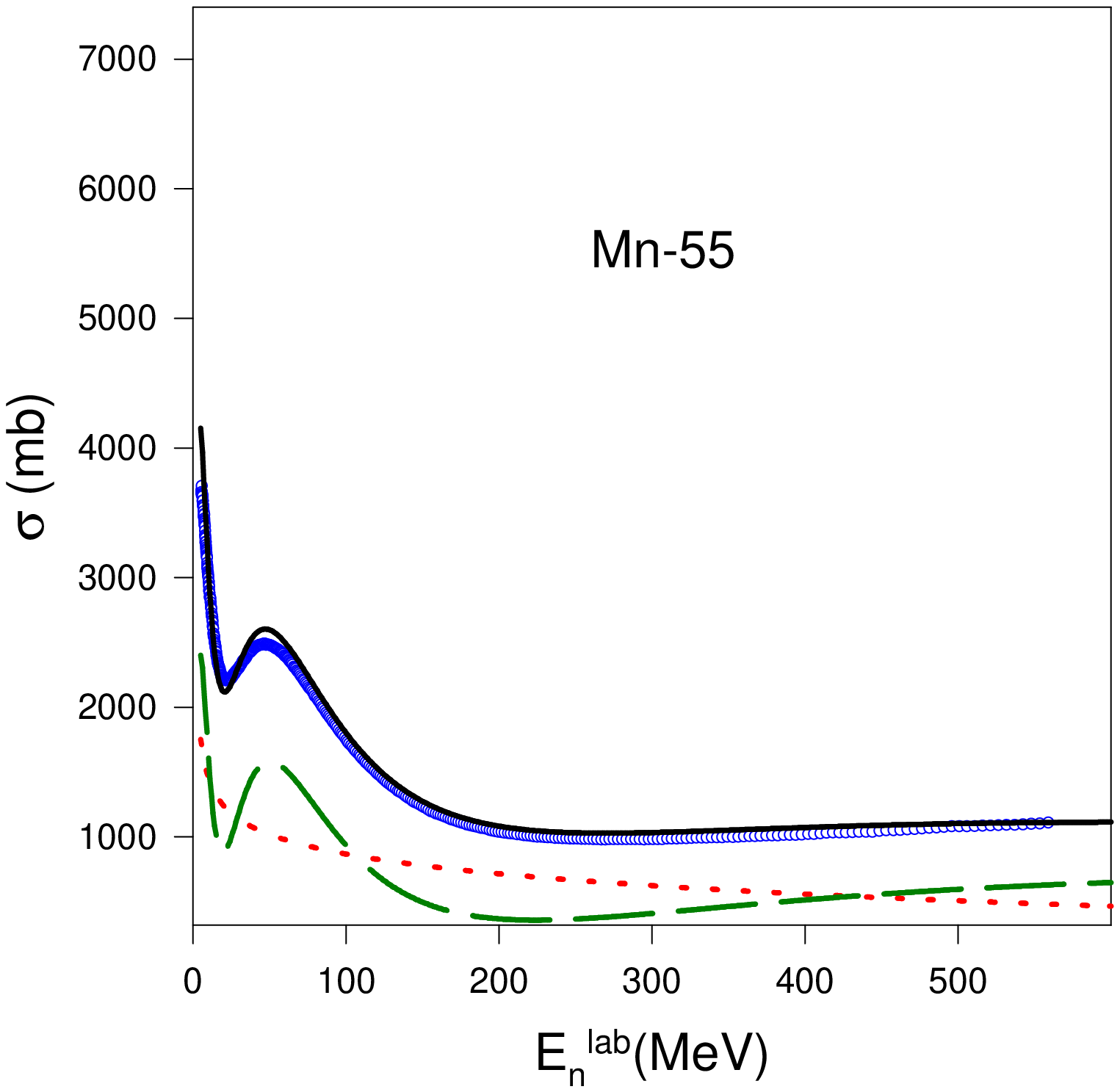,height=10cm,width=10cm}}
\caption{}
\label{fig13}
\end{figure}
\pagebreak

\begin{figure}[h]
\eject\centerline{\epsfig{file=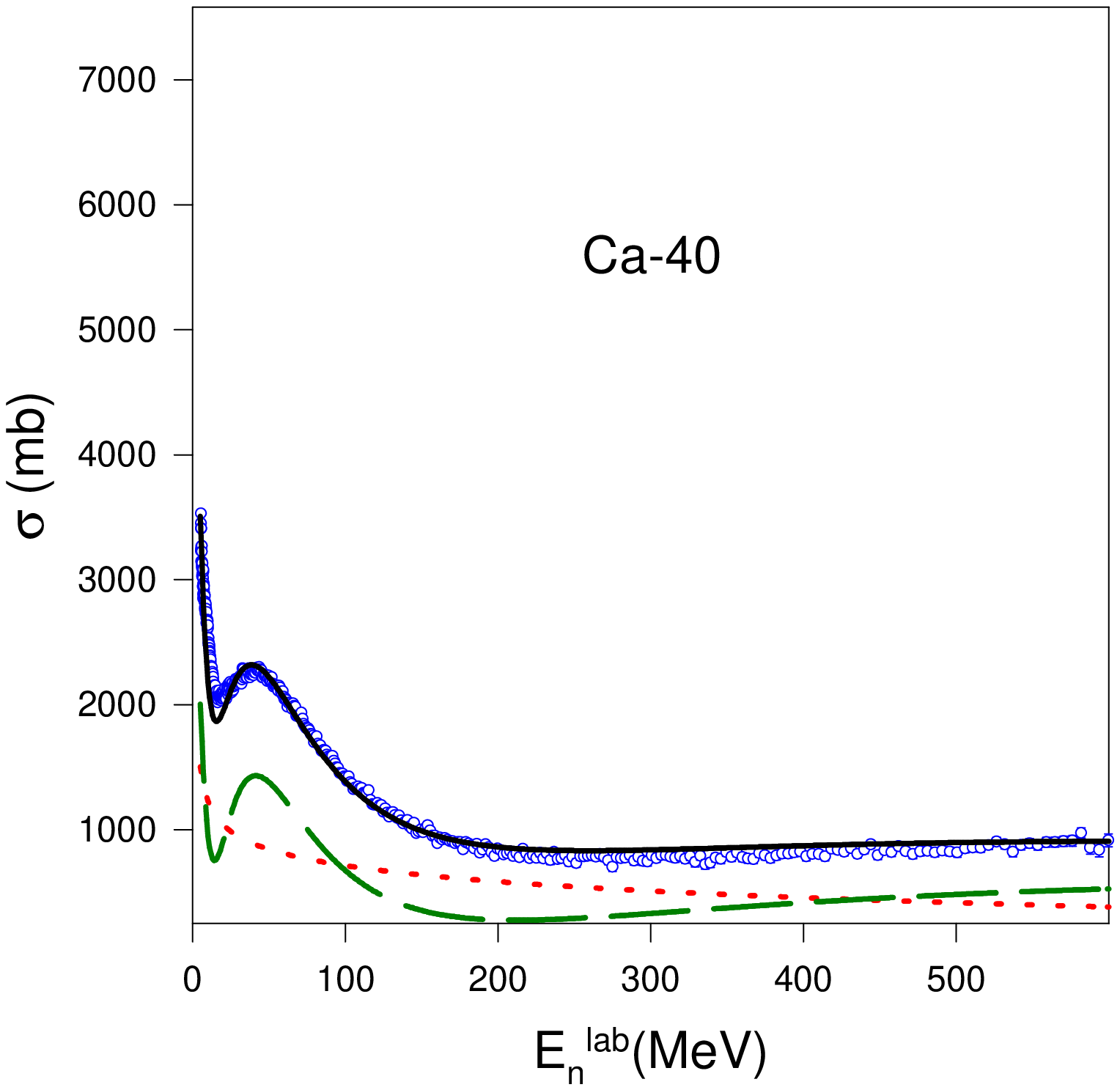,height=10cm,width=10cm}}
\caption{}
\label{fig14}
\end{figure}
\pagebreak

\begin{figure}[h]
\eject\centerline{\epsfig{file=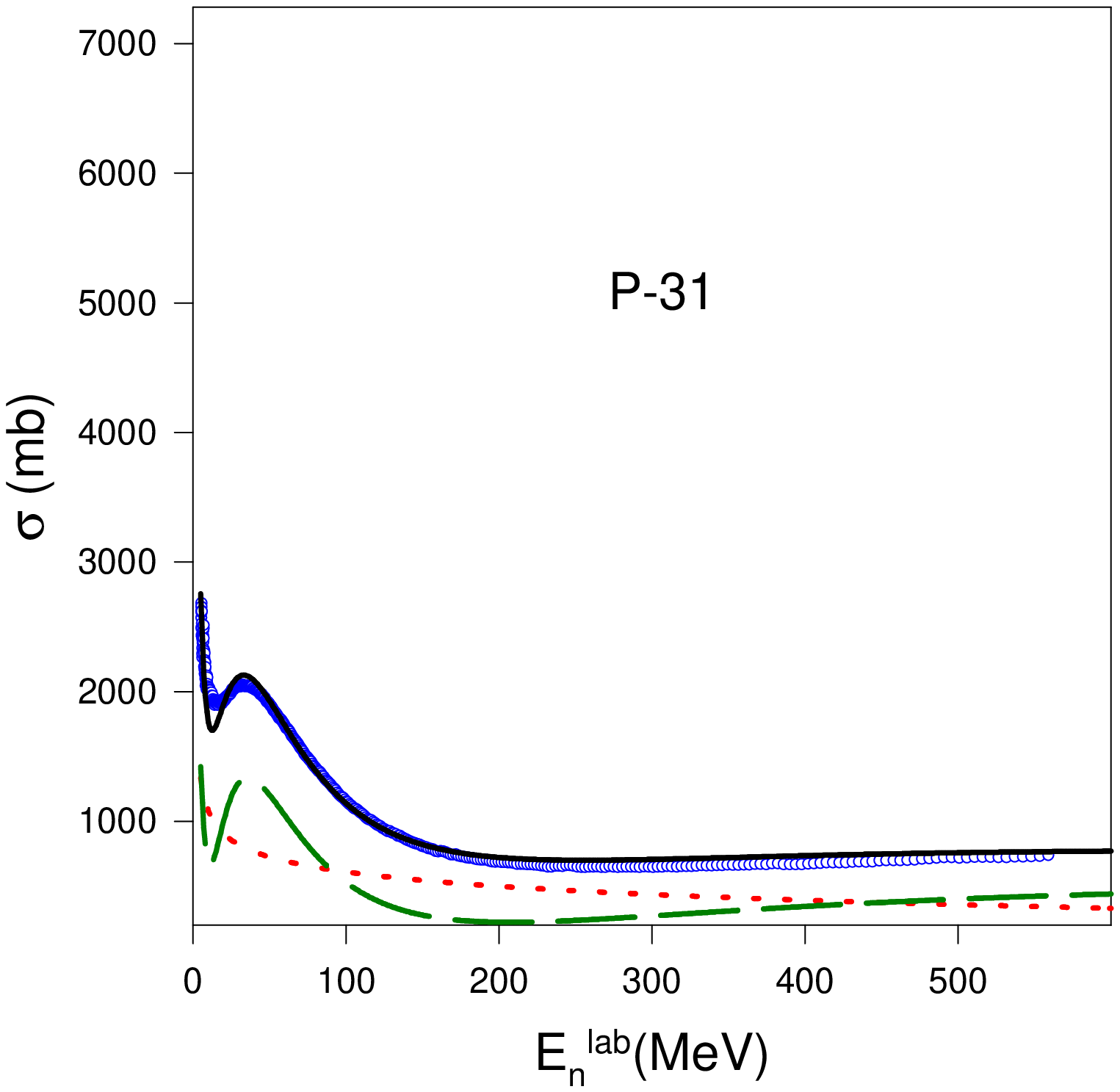,height=10cm,width=10cm}}
\caption{}
\label{fig15}
\end{figure}
\pagebreak

\begin{figure}[h]
\eject\centerline{\epsfig{file=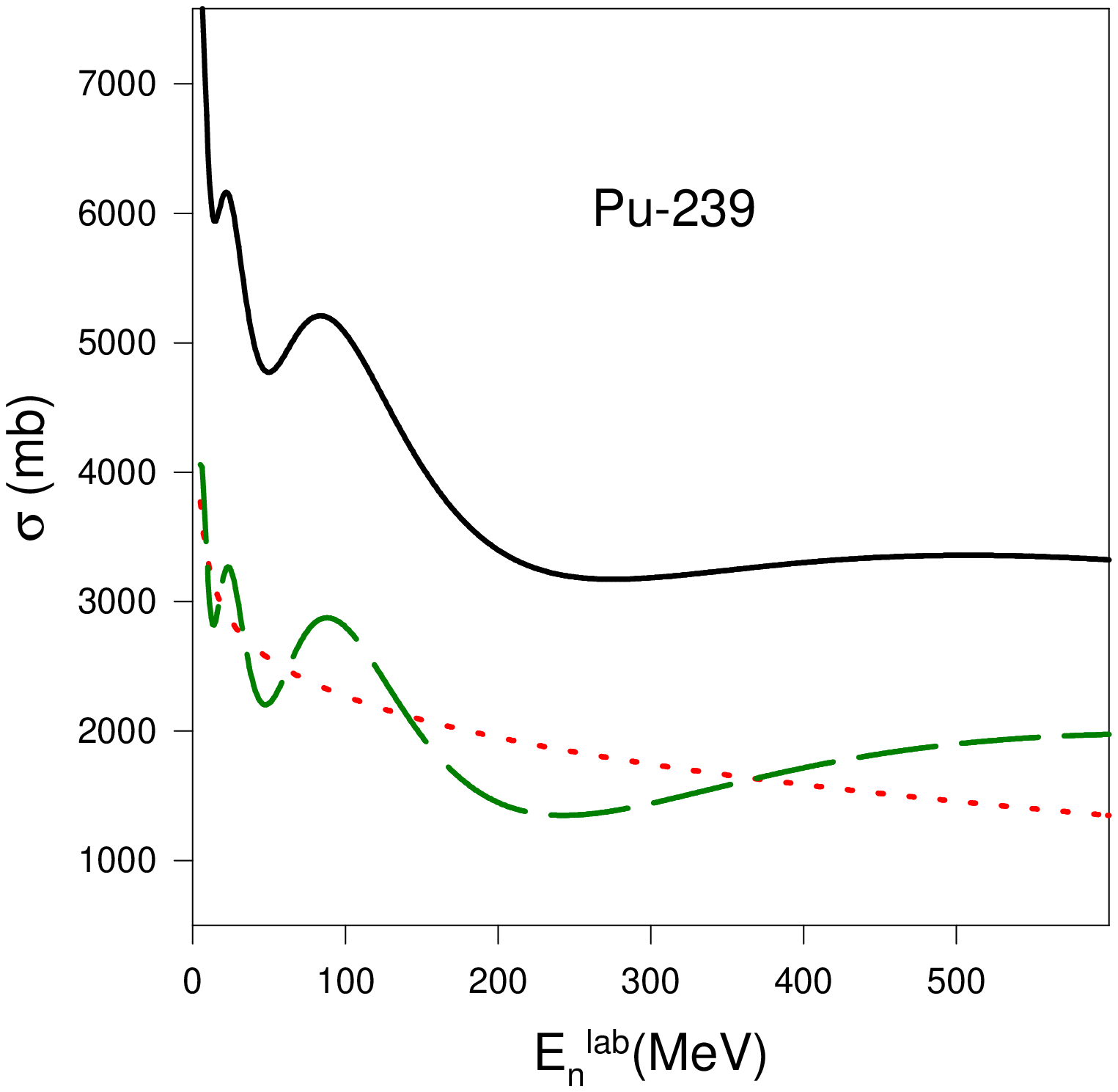,height=10cm,width=10cm}}
\caption{}
\label{fig16}
\end{figure}
\pagebreak
          
\noindent
{\bf Figure captions }

{\bf Fig.1.} The plots of total cross section ($\sigma_{tot}$), the scattering cross section ($\sigma_{sc}$) and the reaction cross section ($\sigma_r$) versus incident neutron energy for $^{238}$U target. The continuous line represents the total cross section, the dashed line represents the scattering cross section, the dotted line represents the reaction cross section and the hollow circles represent the experimental data.

{\bf Fig.2.} Same as Fig.1 but for $^{232}$Th target. 

{\bf Fig.3.} Same as Fig.1 but for $^{209}$Bi target. 

{\bf Fig.4.} Same as Fig.1 but for $^{208}$Pb target. 

{\bf Fig.5.} Same as Fig.1 but for $^{197}$Au target. 

{\bf Fig.6.} Same as Fig.1 but for $^{186}$W target. 

{\bf Fig.7.} Same as Fig.1 but for $^{184}$W target. 

{\bf Fig.8.} Same as Fig.1 but for $^{182}$W target. 

{\bf Fig.9.} Same as Fig.1 but for $^{181}$Ta target. 

{\bf Fig.10.} Same as Fig.1 but for $^{93}$Nb target. 

{\bf Fig.11.} Same as Fig.1 but forr $^{90}$Zr target. 

{\bf Fig.12.} Same as Fig.1 but for $^{59}$Co target. 

{\bf Fig.13.} Same as Fig.1 but for $^{55}$Mn target. 

{\bf Fig.14.} Same as Fig.1 but for $^{40}$Ca target. 

{\bf Fig.15.} Same as Fig.1 but for $^{31}$P target. 

{\bf Fig.16.} The plots of total cross section ($\sigma_{tot}$), the scattering cross section ($\sigma_{sc}$) and the reaction cross section ($\sigma_r$) versus incident neutron energy for $^{239}$Pu target. The continuous line represents the total cross section, the dashed line represents the scattering cross section and the dotted line represents the reaction cross section.
                   
\end{document}